\documentclass[twocolumn]{revtex4}
\usepackage{graphicx}
\newtheorem{criterion}{Criterion}
\begin{document}
\title{On experimental procedures for entanglement verification}
\author{S.J. van Enk$^{1,2}$, N. L\"utkenhaus$^3$, and H.J. Kimble$^{2,4}$}
\affiliation{
$^1$Department of Physics, University of Oregon\\
Oregon Center for Optics and
Institute for Theoretical Science\\
 Eugene, OR 97403\\
$^2$Institute for Quantum Information, California Institute of Technology, Pasadena, CA 91125\\
$^3$Institute of Quantum Computing
and Department of Physics and Astronomy\\
University of Waterloo\\
Waterloo, Ontario, N2L 3G1, Canada\\
$^4$Norman Bridge Laboratory of Physics 12-33\\
California Institute of Technology\\
Pasadena, CA 91125}
\begin{abstract}
We give an overview of different types of entanglement that can be generated in experiments, as well as of various protocols that can be used to verify or quantify entanglement.
We propose several criteria that, we argue, should be applied to experimental entanglement verification procedures.
Explicit examples demonstrate that not following these criteria will tend to result in overestimating the amount of entanglement generated in an experiment
or in infering entanglement when there is none. We distinguish protocols meant to refute or eliminate hidden-variable models
from those meant to verify entanglement. 
\end{abstract}
\maketitle
\section{Introduction}
Entanglement plays a crucial role in quantum information processing protocols 
such as quantum computing, teleportation and quantum key distribution.
For example, although the standard Bennett-Brassard 1984 (BB84) protocol \cite{BB84} for quantum key distribution does not require entanglement, it is equivalent to a different protocol that does use a bipartite entangled state. One necessary (although not sufficient) condition for security of the BB84 protocol is that the effective bipartite state from the equivalent protocol is  demonstrably entangled \cite{curty}.
It is thus reasonable 
to say that verifying experimentally created entanglement is of great importance.
However, there are many subtle issues in applying the theory of entanglement verification to actual experiments. These subtleties
 have occasionally led to controversies and misunderstandings, so that it is probably useful to establish some criteria for discussions of experimental protocols designed to detect, measure, or verify entanglement.
Formulating such criteria is one purpose of the present paper. In addition it provides a survey of the many different methods that have become available in recent years to characterize entanglement operationally. We focus here on bipartite entanglement only.

Depending on which entanglement verification protocol one uses, it may still be true that the entangled states one has generated and verified are not useful for the task one had in mind. Indeed, for most tasks that require entanglement, one also must have a good estimate of the state itself and in which Hilbert space it lives. However,
 most of the protocols we will consider here tell one whether there is entanglement, but not precisely what state one should assign. Some other protocols even tell one how much entanglement one generates, but still do not yield an estimate of the state. Those protocols then must be supplemented by other protocols estimating the state itself.
Furthermore, in addition to having to know what entangled state one generated,
one typically also needs a certain minimum amount of entanglement for it to be useful for a specific task. For example, in order to perform a fault-tolerant quantum computation 
via measurements, along the line of Gottesman and Chuang \cite{gotchuang}, one needs {\em much} more entanglement than for violating a Bell inequality. In the present paper, though, we will be mostly concerned with the simplest nontrivial task of establishing whether there is entanglement or not.
That task may be hard enough in practice.
\subsection{Types of entanglement}
Let us begin with a short overview of the various types of (bipartite) entanglement that can be generated in experiments, and in what sort of physical systems they tend to occur.
We distinguish three main categories that
all refer to genuine entanglement, but the usefulness for quantum information processing protocols varies from one category to the next. 
\begin{enumerate}
\item \underline{{\em A priori} entanglement}: Here one has a source generating many instances (or ``copies'' \footnote{The word ``copy'' does not imply that some unknown quantum state is copied by the source. Rather, the source applies the same classical preparation procedure in each instance.}) of a bipartite state of systems $A$ and $B$, 
\begin{equation}\rho_{AB}=
\rho_{{\rm ent}}.\end{equation}
Here we assume one has performed measurements on a subset of the many ``copies'' that allow one to give an accurate estimate of the state $\rho_{{\rm ent}}$.
Then, depending on the amount of entanglement in $\rho_{{\rm ent}}$, the remaining unmeasured copies can be used for teleportation,
for quantum computation, or for quantum key distribution.
For instance, in a quantum computation the entanglement should be within $\epsilon$ of the maximum
where $\epsilon$ is the threshold for fault-tolerant quantum computation \cite{fault} in order to be useful.
For a quantum repeater the entanglement should typically be within a few percent of the maximum \cite{qurepeat}.
This type of {\em a priori} entanglement can be generated, e.g., in ion traps where two ions can be entangled in a deterministic way \cite{ionent}, or in two-mode squeezed states of light, where continuous-variable entanglement is deterministically created \cite{CVent1,CVent}, or between two atomic ensembles
when analogous continuous-variable degrees of freedom (Stokes parameters) are used \cite{CVentmatter}. 
\item \underline{Heralded entanglement}: Suppose one's source generates
many instances of states of the form
\begin{equation}
\rho_{AB}=p_{{\rm yes}}\rho_{{\rm ent}}\otimes \rho_{{\rm yes}}+
p_{{\rm no}}\rho_{{\rm unent}}\otimes \rho_{{\rm no}}.
\end{equation}
Here $\rho_{{\rm ent}}$ is an entangled state, which one has subjected to many tests so that one has a reliable estimate of that state.
The states $\rho_{{\rm yes}}$ and $\rho_{{\rm no}}$ are (orthogonal) states of an auxiliary system, on which one performs
measurements that tell one whether one is left with the entangled state $\rho_{{\rm ent}}$ or the unentangled (and useless) state
$\rho_{{\rm unent}}$. The probabilities to project onto the entangled or unentangled states are denoted by $p_{{\rm yes}}$ and $p_{{\rm no}}$. 

This type of entanglement is almost as useful and as powerful
as {\em a priori} entanglement, except that one may have to generate many copies before achieving success  if $p_{{\rm yes}}$ is small.
Indeed, one needs to generate on average $1/p_{{\rm yes}}$
copies before one is able to make use of the entanglement. This type of entanglement is generated, e.g., between atomic ensembles \cite{chou}, using the protocol
from Ref.~\cite{dlcz}. Here detection of a single photon that emanates from one of two ensembles {\em heralds} the creation of entanglement between the two ensembles.
\item \underline{{\em A posteriori} entanglement}:
Here one generates many copies of a state
\begin{equation}
\rho_{AB}=(1-P)\rho_0 + P \rho_{{\rm ent}},
\end{equation}
where $P\ll 1$ is a small probability, $\rho_0$ is a state
that one's detection device does not detect, and $\rho_{{\rm ent}}$ is one's desired entangled state 
(e.g., a state close to a maximally entangled Bell state) that is destructively detected by one's devices.
In this case, the amount of entanglement is actually small (of order $P$) but one can still detect it and one can use it for some but not necessarily all protocols that require entanglement. 

In fact, because of the destructive character of one's detection methods, one typically uses (and sometimes describes \cite{kwiat,swapping}) the state $\rho_{AB}$ as if it is close to being maximally entangled, that is, as if it is in fact the state $\rho_{{\rm ent}}$.
But most of the time (with probability $1-P$) one's detectors 
do not register any signal and the protocol for which the entanglement is supposed to be used fails. But upon a positive detection event one did succeed in implementing one's desired protocol {\em a posteriori}.
This type of entanglement arises, e.g., in 
typical down-conversion experiments,
where $\rho_0$ is a state containing no photons, and $\rho_{{\rm ent}}$ contains a pair of photons, one photon for each party. (We neglect here the probability of order $P^2$ to have two pairs of photons, $\rho_{2p}$.
This is a good approximation only when considering a {\em single} copy of the state $\rho_{AB}$, but not when one considers two copies or more. After all, to order $P^2$ one has not only two copies of $\rho_{{\rm ent}}$ but there are also terms corresponding to the case where one has one copy of $\rho_0$ and one of $\rho_{2p}$. See \cite{slb} for examples.) 
 Such entanglement can be useful for generating classical data displaying nonclassical correlations, but not always for generating quantum outputs.
For example, {\em a posteriori} entanglement
 can be used for Bell inequality tests \cite{kwiat,loop1}. The main difference between {\em a posteriori} entanglement and heralded entanglement is that in the latter case the entangled state $\rho_{{\rm ent}}$ exists after one's measurement, ready to be used, while in the former case the desired state  $\rho_{{\rm ent}}$ never exists with a high fidelity (since $P\ll 1$).
Before one's measurement one has $\rho_{AB}$, after one's measurement one has destroyed the state.

On the other hand, one can in principle promote {\em a posteriori} entanglement to heralded entanglement by using {\em two copies} of the state and applying entanglement swapping \cite{swapping}. That is, conditioned on detecting two photons, {\em one from each copy}, and projecting those two photons in a Bell state, one has to a very good approximation another pair of propagating photons that should be close to the desired maximally entangled state. We would not actually agree with the name ``entanglement swapping'' here as that term would imply one starts off with a close to maximally entangled state, whereas one does not. Moreover, in the actual experiment \cite{swapping}
the detection of two photons in the Bell measurement could be due (with roughly 50\% probability)  to two photons from {\em one} mode (arising from the term $\rho_{2p}$ mentioned above). So, with the method of \cite{swapping} one actually produces a roughly equal mixture of a close-to-maximally entangled state of two photons, and a state with no photons in one output and two (unentangled) photons in the other. For a more precise and detailed discussion of this issue, see \cite{kokb}.
\end{enumerate}
\subsection{Overview of entanglement verification protocols}
We can also distinguish various entanglement verification protocols. We briefly discuss them here in no particular order (except for being treated in this order, much more extensively, in the next Sections), and also indicate what type of not-so-subtle errors have been made in applying these verification protocols (more subtle errors are discussed later):
\begin{enumerate}
\item \underline{Teleportation}: Here one tries to use entangled states to teleport some ensemble of quantum states. If the average teleportation fidelity is sufficiently high, one has proven the existence of entanglement (for more details see Section \ref{telep}). 

Just as there are different types of entanglement, there are different types of teleportation. With {\em a priori} entanglement one can in principle perform deterministic and unconditional teleportation, provided one can perform a Bell measurement in a deterministic way. That is, 
one can guarantee that the quantum state of a given system will be teleported with high fidelity and high probability.
This type of teleportation has been achieved both with ions in an ion trap \cite{wine,blatt} as well as with two-mode squeezed states \cite{CVent,CVentm1,CVentm2}.

With heralded entanglement one can do the same (after a successful heralding event), but with {\em a posteriori} entanglement one can only perform conditional teleportation. That is, the success of the teleportation protocol cannot be guaranteed in advance. The success is conditioned on the positive outcomes of certain measurements, including the Bell measurement, that are part of the teleportation protocol itself. For example, in the pioneering experiment of Ref.~\cite{bouw} successful teleportation could be inferred  only after a succesful Bell measurement, {\em and} after detecting and destroying the teleported state. We may call this conditional  {\em a posteriori} teleportation. The improved version \cite{slb} of teleportation with {\em a posteriori} entanglement of Ref.~\cite{un} no longer relies on having to detect and destroy the teleported state. That type of teleportation is still conditional, success being conditioned on the positive outcome of the Bell measurement. 

In the context of the present paper it is important to note that {\em all} these versions of teleportation (conditional or unconditional, {\em a posteriori} or {\em a priori}) can be used as valid entanglement verification tests.
In order to quantify the amount of entanglement generated one does have to take into account in what fraction of the attempts successful teleportation was achieved. But a sufficiently high fidelity
in the successful cases does demonstrate the presence of entanglement.
\item \underline{Bell-CHSH inequality tests}: Here one subjects the entangled states to measurements whose correlations may turn out to be so strong that they cannot be explained by a local hidden-variable model \cite{bell,chsh}. This, in turn, implies entanglement, as 
any unentangled state could act as a local hidden-variable model.
Null results, where no signal was detected, may be taken care of by a ``no-enhancement'' or a ``random-sampling'' assumption (see Section \ref{bell} for more details).  Thus, {\em any} of the three types of entanglement described above can be tested and verified by a Bell-CHSH \footnote{Also sometimes called CSHS-inequality.} inequality test.

The Bell-CHSH test is very powerful in that it makes no assumptions about Hilbert spaces involved, nor about what one's measurement devices actually detect.
On the other hand, the amount of the violation of such an equality
by itself tells one nothing quantitative about the amount of entanglement generated in the experiment. 
For example, in the case of {\em a posteriori} entanglement there is a clear difference between $\rho_{AB}$ and $\rho_{{\rm ent}}$.  The violation of Bell's inequality tells one only about the nonzero entanglement in $\rho_{{\rm ent}}$ but not about the entanglement in the state actually generated, $\rho_{AB}$.
Including the fraction of null results does give one the correct order of magnitude of entanglement in $\rho_{AB}$, but typically one does not include these in statements about entanglement \cite{kwiat2}. 
\item \underline{Tomography}: One performs full tomography \footnote{We mean by ``tomography'' any reliable method of quantum state estimation. For an analysis of more and less reliable quantum state estimation methods, see \cite{rbk}.} on a quantum state and from that infers, by calculation, how much entanglement one has. This seems straightforward,
but one has to be careful here. In the case of {\em a priori} entanglement the state one performs tomography on, $\rho_{{\rm ent}}$, actually is the state generated in one's experiment, $\rho_{AB}$. In the case of {\em a posteriori} entanglement one typically performs tomography not on the full state, but only on the part that is detected, $\rho_{{\rm ent}}$. {\em That} state $\rho_{{\rm ent}}$ never exists independently in one's experiment, and conclusions based on that state overestimate the amount of entanglement in $\rho_{AB}$ by a factor of $1/P$.
For example, although a graph as in, e.g., \cite{kwiat} on tomography of a state generated by down conversion, looks similar to a graph in a tomography experiment on an entangled state between two ions \cite{tomoion}, only the latter experiment performed tomography of the full quantum state.
\item \underline{Entanglement witnesses}: One measures a particular observable whose expectation value is positive for any unentangled state \cite{ew}. If one finds a negative value, one must, therefore, have entanglement.
Null measurements can be counted as contributing ``zero'' to the entanglement witness, and, therefore, do not affect the sign of the entanglement witness. Thus, measuring an entanglement witness is a valid test for all three types of entanglement discussed above.
\item \underline{Direct measurement of entanglement}: Here one measures certain quantities that not only tell one there is entanglement, but also how much.
This requires one to have multiple copies of the same state \cite{horodecki,horodecki2,carteret}.
However, what one means precisely by ``multiple copies'' or ``identical copies'' is not trivial. Placing too much trust on having identical copies without testing this first can lead one to wrong predictions about entanglement. A simple example is \cite{walborn} where measurements are performed on only one part of a bipartite state, and yet maximal entanglement is concluded to exist. Here one relies on the strong additional assumption that one generated two identical {\em pure} states. For more details, see Section \ref{direct}.
\item \underline{Consistency with entanglement}:
Here one's (ideal) theory tells one the experiment should, if all goes well, generate some entangled state $\rho_{{\rm ent}}$. One performs certain measurements and checks that one's measurements are consistent with the assumption of the entangled state $\rho_{{\rm ent}}$. However, in general
one cannot infer the existence of entanglement from these measurements and the measurement results will typically be consistent with a classically correlated but unentangled state as well. A recent example of this procedure is \cite{haroche} (see also \cite{kuzmich}).
\end{enumerate}
In the remainder of this paper we will be concerned with the correct entanglement verification protocols 1--5 only, and we exclude protocol 6 (it does feature in Section \ref{direct} because a particular incorrect application of protocol 5 is similar to protocol 6). Nevertheless, even correct protocols must be applied carefully, and discussing the most common mistakes in applying correct protocols is one point of the present paper. Those errors tend to be more subtle than the ones we just mentioned in this brief overview.
\subsection{The different roles played by the experimentalist}
It will be convenient in the following discussions to introduce five characters playing different roles
in experiments that generate and verify entanglement. First there are the usual personae {\em Alice} and {\em Bob} who claim to have generated a bipartite entangled state shared between the two of them. Alice and Bob are in separate locations $A$ and $B$. 

Then there is {\em Quinten} \footnote{Quinten is not a misspelling of the name Quentin,
but the name of a character from a Dutch novel ``De ontdekking van de hemel'' (translated as ``The discovery of heaven''), by H. Mulisch.} who wishes to verify that Alice and Bob have indeed generated entanglement. We assume Alice and Bob hand over their entangled state to Quinten, who then subjects that state to his favorite test (and this is repeated many times).
Quinten believes in quantum mechanics but does not trust Alice and Bob.
 
Quinten should be distinguished from another verifier, {\em Victor}, who was introduced some time ago in the literature \cite{CVent} in the context of teleportation protocols. Victor lets Alice and Bob teleport a state that he hands over to Alice. The state is known to Victor, but not to Alice and Bob. Victor checks whether the output state on Bob's side at the end of the teleportation protocol is sufficiently close to his original input state to warrant the conclusion Alice and Bob must have made use of entanglement (more details are given in Section \ref{telep}). In contrast to Victor, Quinten performs all tests himself. In particular, if Quinten uses teleportation as a means to verify entanglement, he himself would try to teleport some arbitrary state using the entangled state handed to him by Alice and Bob.

Finally, there is {\em Rhiannon} \footnote{Rhiannon is a figure from Welsh mythology whose name starts with an R.}, the realist, who, just like Quinten, mistrusts Alice and Bob and performs her own measurements on states handed over to her by Alice and Bob. But unlike Quinten, she does not believe in quantum mechanics and tries to construct a local (``realistic'') hidden-variable model that describes her measurement results. She does not accept any quantum-mechanical descriptions of (measurement or other) devices and interprets results of measurements only in terms of the classical settings of those devices 
and in terms of the different ``clicks'' her measurement devices produce. 

In the following we will be mostly interested in verification procedures that could convince Quinten that Alice and Bob share entanglement. This preference reflects our own belief in the validity of quantum mechanics. 
Quinten's protocols do not test quantum mechanics; they merely test the state handed over to him by Alice and Bob. 
Such protocols do not necessarily convince Rhiannon of anything, as she will not accept any of the inferences made by Quinten that depend on the validity of quantum mechanics.

On the other hand, protocols designed to convince Rhiannon that Alice and Bob's measurement results cannot be explained by a local hidden-variable model, are perfectly acceptable as a demonstration for Quinten that Alice and Bob made essential use of entanglement. This is simply because any unentangled state can itself be used as a local realistic hidden variable model. Moreover, such an experiment may convince Quinten of entanglement even if it fails in Rhiannon's eyes. Indeed, the presence of ``loopholes'' in Bell-inequality tests or CHSH-inequality tests \cite{santos}
make the test unacceptable for Rhiannon, but Quinten is willing to make more assumptions about the experiment (see below for examples). Thus he may accept the violation of a Bell or CHSH inequality as verifying the presence of entanglement even if loopholes are in fact present.
An explicit example demonstrating that using more assumptions  allows one to relax the conditions under which entanglement is verified is provided by Ref.~\cite{seevinck}, where
a more relaxed Bell inequality is derived under additional assumptions
about the (quantum) description of the measurements performed.
It is important to distinguish these assumptions about quantum mechanics made by Quinten from other assumptions that would boil down to trusting Alice and Bob. The latter, we argue, should {\em not} be allowed in entanglement-verification experiments. 

Thus we propose that {\em every entanglement-verification protocol follow Quinten's criteria and assumptions.}
In the next section we will formulate some Criteria that Quinten should use (mostly arising because of his mistrusting Alice and Bob). In the subsequent sections we will discuss examples of assumptions often made in experiments that do not violate these Criteria, examples of assumptions that do violate them, and what the consequences are of such violations. 
\subsection{What is entanglement verification anyway?}
We conclude this Introduction by defining what exactly we mean when we say we verified that a source creates bi-partite entangled states. This question is meaningful only in the situation that the source emits a long sequence of repeated signals. It is not clear in general that we can describe this situation via density matrices for the individual signals. However, in special cases this can in fact be done. 

The answer to the question what state to assign under which circumstances  has been given in \cite{finetti} for the following case, relevant to our discussion. Assume Alice and Bob generate $\tilde{N}$ copies of their bipartite state.
Then consider the case in which Quinten tests some randomly chosen subset of $N<\tilde{N}$ copies.
Suppose he concludes that the sequence of $N$ copies is exchangeable, i.e., he assigns the same overall state to
any permutation of these $N$ systems. Furthermore, 
assume Quinten considers this exchangeability to hold true for an extended sequence of $N+1$ states created by adding one more copy.
This assumption is called extendability.
With both conditions of exchangeability {\em and} extendability satisfied he then should assign a state of the form
\begin{equation} \label{dF}
	\rho^{(N)}=\int {\rm d}\rho P(\rho) \rho^{\otimes N},
\end{equation}
with ${\rm d}\rho P(\rho)$ some probability measure, to his sequence of $N$ copies.
This form is called the De Finetti form \cite{finetti}. Note that this form
is more general than the simple relation 
\begin{equation} \label{simpledF}
\rho^{(N)}=\rho_0^{\otimes N},
\end{equation}
 for $N$ copies, where $\rho_0$ is some fixed {\em known} density matrix.
 Indeed, the latter form is a special case of the De Finetti form, namely when Quinten has narrowed down his probability assignment $P(\rho)$ to a delta function $\delta(\rho-\rho_0)$. For instance, he can achieve this by performing full quantum state tomography. 
 
More generally, Quinten's measurement statistics restrict the form of those $\rho$ that contribute non-negligibly in (\ref{dF}), and verifying entanglement means to verify that {\em all} contributing $\rho$'s are entangled. That answers the question posed in the title of this subsection.
 
 Often one uses the statement that the density matrix $\rho^{(N)}$ of $N$ systems is ``of the form $\rho^{(N)}=\rho^{\otimes N}$'' to imply that $\rho$ is {\em not} known. In that case, one could or perhaps should use the more complicated form (\ref{dF}) to make the unknown character of $\rho$  explicit. With this interpretation in mind,
 it is permissible to assume the form $\rho^{(N)}=\rho^{\otimes N}$ if exchangeability and extendability both hold. This is the case, for example, if all signals are being measured independently, so that the exchangeability and extendability properties apply to the classical data (i.e. the measurement results). 

Clearly, verifying entanglement in this sense can be done only with a sufficient amount of measurement statistics. The form (\ref{dF}) is, in fact, valid asymptotically, for $N\rightarrow\infty$, and neglects terms that vanish in the limit $N \to \infty$. The question arises how fast these terms decrease with increasing $N$. The discussions of those details is beyond the scope of the present paper, and for more discussion on that topic, see Ref.~\cite{Rennerthesis,renner}. But it is important to keep this issue in mind when making the statement that one's source emits entangled states. 
\section{Criteria for experimental entanglement verification}\label{criterion}
Here we propose five Criteria that any entanglement-verification procedure should satisfy.
We illustrate how violating these Criteria tends to lead to overestimating the amount of entanglement in the entangled state generated, or to concluding that there is entanglement when there is in fact none. In this Section we use one very simple example to illustrate some of the Criteria. Suppose the bipartite state of two qubits one generates is of the form
\begin{eqnarray}\label{example}
	\rho&=&\int\frac{{\rm d}\phi}{2\pi} |\Psi(\phi)\rangle \langle \Psi(\phi)|;
	\nonumber\\
|\Psi(\phi)\rangle&=&	\frac{(|0\rangle|1\rangle+\exp(i\phi)|1\rangle|0\rangle)}{\sqrt{2}}.
\end{eqnarray}
This state is a mixture over a phase $\phi$ of maximally entangled states.
Yet the mixed state has no entanglement, which becomes clear after performing the integral over $\phi$.
Violating the Criteria given below, however, may lead one to conclude incorrectly there {\em is} entanglement.
In later Sections we will give more complicated examples from actual experiments violating the Criteria in more subtle but equally disallowed ways. We will also display examples in those Sections to explain why sometimes complicated procedures are required even if they may seem unneccesary or overcautious at first sight. 

Any experimental result must be interpreted before it can be checked against a theory.
Whereas Rhiannon only translates the classical settings of devices and classical outcomes of measurement into symbols,  
Quinten translates measurements made by him
into mathematical expressions corresponding to von Neumann measurements or, more generally, to Positive-Operator-Valued-Measures (POVMs).
Now in an actual experiment the roles of both Alice and Bob on the one hand, and that of Quinten on the other, are 
played by the same experimentalist. Thus it is an easy mistake to accept quantum-mechanical descriptions of {\em all} operations as valid. However, in the scenario sketched in the Introduction, Quinten does not trust Alice and Bob. This means that
quantum-mechanical descriptions of operations performed in the {\em preparation} procedure (the part of an experiment assigned to Alice and Bob) should {\em not} be taken for granted. On the other hand, measurements done during the verification stage (Quinten's measurements) can be  
trusted, although, of course, unjustified assumptions about Quinten's measurements should be avoided as well. 

Almost all over-optimistic statements about entanglement generated in actual experiments put too much trust in the preparation stage.
Thus the first Criterion may well be the most important:
\begin{criterion}\label{state}
	No assumption should be made about the form of the quantum state generated in the experiment.
\end{criterion}
For example, suppose an experiment generates a state of the form (\ref{example}). Now the intention may well have been to have full control over the phase $\phi$. However, one should not simply assume one succeeded in doing that. In the case of (\ref{example}) one would prematurely conclude one generated a maximally entangled state by assuming a particular value of the phase, although there is no entanglement.
Thus, we argue, verifying that one has control over the phase $\phi$ should be part of the entanglement verification protocol.
More interesting examples of violations of Criterion 1 are found in Sections \ref{bell} and \ref{direct}.

A special case of this Criterion, but probably worth stating explicitly, is
\begin{criterion}\label{symm}
	No assumption should be made about the symmetries of the quantum state generated in the experiment.
\end{criterion}
Using the same example Eq.~(\ref{example}), one may well decide that all phases are equivalent in one's experiment ``by symmetry'', and therefore one decides that one might as well set $\phi$ to zero {\em by fiat}. This would lead, again, to the wrong conclusion that one generated a maximally entangled state, if one actually generated (\ref{example}). More interesting examples are in Section \ref{telep}.

Another special case of Criterion \ref{state} worth stating explicitly refers to the form of the state of {\em multiple} copies
of a quantum system (see also \cite{renner}). 
\begin{criterion}\label{multi}
	One should not assume $N$ copies of the state generated in an experiment to be
	in a joint state of the form $\rho^{(N)}=\int d\rho P(\rho)\rho^{\otimes N}$ unless the verification measurements demonstrate exchangeability and extendability of the sequence of  the $N$ copies.
\end{criterion}
This, of course, refers to the discussion about the De Finetti form (\ref{dF}).
Indeed, it is important to emphasize this: the assumptions of Quinten about exchangeability and extendability
should {\em follow from his measurements}, not from his trusting Alice's and Bob's actions.
And so his measurements should be done as follows:
First, if he perform various different measurements $M_i$, he should choose a random order for those measurements on his copies to ensure he can apply exchangeability. Second, in order to warrant the extendability assumption, he should be able to delete
randomly chosen subsets of his data and still reach the same conclusion about his state assignment (or about entanglement).
These two Criteria are sufficient for him provided all his measurements $M_i$ are performed on {\em single} copies. In the case that some or all of his measurements are performed jointly on groups of two or more copies, then he should in addition choose those groups of copies randomly.
Examples of where these procedures are necessary will be given in Section \ref{direct}, about direct measurements of entanglement.

Let us note explicitly that Quinten will assign correctly the form $\rho^{(N)}=\int d\rho P(\rho)\rho^{\otimes N}$ to verify entanglement even in cases where one might normally be inclined to assign a different form. For instance, a phase-diffusing laser emitting light pulses in subsequent time slots
can be described by a quantum state of a slightly different but related form \cite{vEF}, that takes explicitly into account that different portions of the laser light within a coherence length of the laser differ only slightly in phase, whereas two portions farther apart in time may have very different phases. That description, however, makes use of extra knowledge about how a laser actually works. Quinten, however, will trust only his measurement results and should make no assumptions about how Alice's and Bob's lasers work.

For our next Criterion, we argue that the way the verification test is performed in the experiment should not depend on
how the actual state is generated. If the same state can be generated in different ways, then Quinten's verification procedure should not depend on which of the possible methods was used by Alice and Bob. Thus, we have another Criterion:
\begin{criterion}\label{independence}
	Entanglement verification should be independent of the entanglement generation procedure, except for the sharing of stable classical resources.
\end{criterion}
For example, suppose again (\ref{example}) is generated. Assume that the reason for integrating over all $\phi$
is that one does not have control over the phase from run to run because some optical path length is unstable. However, suppose that within one run one can be quite sure that the verification process still can make use of almost the same optical path length, whatever it is. In that case the verification measurements from run to run would give the wrong impression that phase {\em is} under control, and again one would overestimate the entanglement in the state (\ref{example}). This same example is discussed further in Section \ref{tomo} for an actual experiment. 

We do allow here the sharing of {\em classical} resources between Alice and Bob on the one hand, and Quinten on the other.
Such resources may act as reference frames (we do allow Quinten to use the same fixed stars as reference frame) or placeholders (we do allow Quinten to use the same ion trap in which entangled ions are stored, or the same optical table, or indeed the same lab.)

A critical case on the borderline between classical and quantum resources is the sharing of lasers (used a phase reference, for example) between Alice and Bob on the one hand, and Quinten on the other. In principle, Quinten should use his own laser to verify entanglement. After all, he should not trust Alice and Bob to have used a stable laser.  However, suppose Alice and Bob use a laser to generate entanglement whose phase drifts over a characteristic time scale $\Delta t$. Suppose further that the time $\Delta t$ is sufficiently long to in principle verify the stability of that laser with respect to some standard phase reference to some relevant accuracy.
As a shortcut Quinten could use, in such a case, the {\em same} laser to verify entanglement, instead of phase locking his own second stable laser to it and then using the second laser to perform the verification.
Thus, although using two independent lasers is safe and correct, we would argue that it is permissible sometimes (for the sake of practicality) to use the same laser during generation and verification, but {\em only if that laser is sufficiently stable}, namely on a time scale sufficiently long to measure the phase of the laser relative to some phase standard. Of course, it is up to Quinten to verify this. Quinten does not just assume anything about the stability of Alice's and Bob's laser, he does have to subject it to his own independent test.
 
Finally, we formulate a Criterion about postselection, which is well-known to cause troubles
for Bell inequality tests \cite{pearle}. This Criterion does not refer to Alice's and Bob's procedures but to Quinten's analysis of his own measurement results.
\begin{criterion}\label{postselection}
	Apply postselection only if it can be simulated by local filtering before and independent of the verification measurements.
\end{criterion}
Here, ``local'' refers to operations that are done separately on Alice's side and Bob's side; it excludes nonlocal operations acting jointly on Alice's and Bob's quantum systems.
That is, Quinten is allowed to apply certain types of postselection, but only if
the subensemble he selects is the same as the subensemble that would be selected if Alice and Bob had applied 
local (filtering) operations \cite{filtering} in their preparation {\em before Quinten's measurements}. That filtering then should in particular be independent of both the choice and the outcomes of Quinten's measurements.
The reason for all this is as follows: A local operation cannot increase the average amount of entanglement. That is,
given a state $\rho_{AB}$ that is generated in some experiment, one has
\begin{equation}
	E(\rho_{AB})\geq p_{{\rm pass}}E(\rho_{{\rm pass}})+p_{{\rm fail}}E(\rho_{{\rm fail}}),
\end{equation}
where $E(.)$ is one's favorite measure of entanglement,
$p_{{\rm pass}}$ is the probability for the local filtering operation to succeed, and $\rho_{{\rm pass}}$ is then the density matrix of the subensemble selected by Quinten. $\rho_{{\rm fail}}$ is the subensemble failing the filtering, and hence thrown out by Quinten.

Postselection applied in this way cannot lead one to believe there is entanglement where there is none. But one can certainly still overestimate the entanglement one generated by misidentifying $\rho_{{\rm pass}}$ with the ensemble $\rho_{AB}$ actually generated in the experiment. This is an error we mentioned before,  in the context of {\em a posteriori} entanglement.
\section{Quantum teleportation}\label{telep}
One way for Quinten to test whether Alice and Bob generated entanglement, is for him to try to use the purported entanglement for teleportation.  Of course, teleportation achieves more (and was designed to achieve more) than merely verifying entanglement, but here we are interested in teleportation only as a means of verifying entanglement. In particular, when one wishes to use teleportation for a quantum repeater \cite{qurepeat} or for quantum computation \cite{gotchuang} one will need more stringent criteria on the fidelity of teleportation than we require here for our limited purposes.
We consider teleportation of both qubits
and continuous-variable (CV) states, i.e. states of bosonic modes.

If Alice and Bob claim to be able to generate 2-qubit entangled states, Quinten may try to verify this by teleporting a qubit prepared in an arbitrary state from Alice's location $A$ to Bob's location $B$.
If he finds he can teleport randomly chosen qubit states with a ``sufficiently high fidelity'', then he can be confident
Alice and Bob prepared a state that is sufficiently close to an ideal Bell state to warrant the conclusion
the state has nonzero entanglement. Similarly, in the CV case, he may try to teleport
an arbitrary state of a given bosonic mode from $A$ to $B$. Again, if the fidelity Quinten finds is sufficiently high, he concludes
Alice and Bob did generate a CV-entangled state. 

Now before discussing in more detail what ``sufficiently high fidelity''
really means, let us compare Quinten's protocol with
a related but different verification protocol, namely one that verifies whether Alice and Bob can do teleportation themselves. This is usually checked by Victor who hands a randomly chosen qubit state over to Alice, who then teleports it to Bob, after which Victor checks, again, the fidelity of the state teleported with respect to the known (to him) original. There is a distinction we can make between the verification protocols of Quinten and Victor.
The difference concerns the use of a shared reference frame between Alice and Bob. In some cases (in fact, this applies more to the continuous-variable case), the shared reference frame may be considered an additional quantum channel \cite{refframes}. While we actually do not agree with this point of view \cite{vEF2,vEF,ref,ref2}, Victor may not be happy about Alice and Bob sharing such a resource as it may seem Alice can cheat by sending Bob directly the state she's received from Victor.
However, this can in principle be circumvented by letting Alice and Bob establish the resource prior to Victor handing over his qubits to Alice. On the other hand, it should be clear that Quinten is allowed to use whatever reference frame he needs to establish an isomorphism between Alice's and Bob's qubit Hilbert spaces \cite{jmod}. So in the present context there is no problem about sharing reference frames during teleportation.
 
Quinten (and similarly, Victor), uses the following procedure: first he chooses some {\em test ensemble} of pure states and associated probabilities
\begin{equation}\label{ensemble}
{\rm test\, ensemble}=\{(|\psi_i\rangle,p_i), i=1\ldots N\}
\end{equation}
of either qubit states or CV states to be teleported.
These states are to be tested in some random order.
Then he calculates how well he could reproduce the state on Bob's side on average by simply
measuring the input state and generating a new state in $B$ dependent on the measurement outcome: in quantum key distribution (QKD) this procedure would correspond to an ``intercept-resend'' attack.
The average fidelity $F$ he finds using the state Alice and Bob provided should be larger than the optimum value
$\tilde{F}$ (given the test ensemble) the intercept-resend attack could produce, because with a separable state one would not be able to do better than that particular limit. 

Now $\tilde{F}$ does depend on the
ensemble chosen by Quinten. The obvious choice would be to use the uniform ensemble over all possible
states. In the qubit case this gives the result that $\tilde{F}=2/3$ \cite{F2}. But Quinten cannot possibly test {\em all} possible states, and a smaller set of test states will in general lead to a higher value of $\tilde{F}$.
Fortunately, one can show \cite{quantumness} for the qubit case that there are simple ensembles consisting of four or six states that lead to the same optimum fidelity of 2/3: one set is the set 
\begin{equation}T\equiv\{(|T_i\rangle,p_i=1/4), i=1\ldots 4\}\end{equation}
 of four tetrahedral states on the Bloch sphere. The other is the set 
 \begin{equation}M\equiv\{(|M_i\rangle,p_i=1/6), i=1\ldots 6\}\end{equation}
  of the six mutually unbiased basis states. Such sets then are eminent candidates to use for verification of entanglement by teleportation \cite{wine,blatt}. 

One may wonder at this point, though, why is it not sufficient to check just one state or perhaps two states? After all, Quinten knows he is {\em not} performing the intercept-resend method, so why should he pretend he has to beat that particular limit? One answer is that in the case the state generated by Alice and Bob is in fact separable, Quinten's procedure can be interpreted as an intercept-resend protocol.
The other answer is, Quinten cannot assume that the fidelity for one or two particular states is representative for the fidelity of the whole ensemble. Indeed, there may well be an asymmetry in the state generated by Alice and Bob that would lead to teleporting certain states much more reliably than others. Quinten would be violating Criterion \ref{symm} by assuming otherwise. 
The only guarantee Quinten needs is that choosing the two ensembles mentioned above {\em cannot} lead to an average fidelity larger than 2/3 if he has a separable state, independent of assumptions about the entangled state generated by Alice and Bob. On the other hand, it is still true he may choose a smaller or different set of test states, but then the fidelity to beat will in general be larger than 2/3.

Similarly, suppose it is obvious from the experimental arrangements that the
teleportation fidelity of any state of the form $|0\rangle+\exp(i\phi)|1\rangle$ is independent of the phase $\phi$.
Can't Quinten then make a shortcut and test only one state out of the ensemble $M$ of that form (that ensemble contains four such states, after all) and use $\tilde{F}=2/3$ as threshold? The answer is again negative, as it would violate Criterion \ref{symm}: although Quinten's setup may well be symmetric under phase shifts, he cannot assume the state generated by Alice and Bob has the same symmetry. For example, in \cite{blatt} a teleportation experiment 
is discussed where only four out of six mutually unbiased basis states were tested. In principle that is not sufficient, although the actual experiment may well have possessed the desired symmetry. There is indeed an intercept-resend attack that reproduces all four mutually unbiased basis states tested in \cite{blatt} with a fidelity of $\tilde{F}=0.77$ by using appropriately constructed POVMs \cite{blume}, beating
the experimental teleportation fidelity of approximately $F\approx 0.75$.
So the simple correct thing to do, independent of one's assumptions about how the experiment works in detail, is to use all four tetrahedral states $T$ or all six mutually unbiased basis states $M$ to estimate the teleportation fidelity and try and beat $\tilde{F}=2/3$.

In the CV case choosing an arbitrarily large set of states ($N\rightarrow\infty$ in (\ref{ensemble})) to be tested for teleportation would lead to an optimum intercept-and-resend fidelity $\tilde{F}=0$, owing to the infinite dimensionality of the relevant Hilbert space. And so it is true that Quinten's task is easy, in principle, to verify CV entanglement. 
Unfortunately, though, arbitrary states are typically not Gaussian (described by a Gaussian Wigner function) and nonGaussian states are in general much harder to generate for Quinten. What has been used as a test ensemble instead is to take a particular subset of Gaussian states, namely the set of coherent states. Then a fidelity
of $\tilde{F}=1/2$ can be reached by the intercept-resend attack \cite{F}. There have been other proposed tests \cite{grangier}, taking the same set of coherent states, but using different criteria. Those criteria, however, do not check for entanglement but for something stronger \cite{grangierreply}. For instance, one may wish to
eliminate hidden-variable models for the teleportation protocol. But in order to convince Quinten no such strong measures are needed. In fact, in the subsection below we will see that Rhiannon will not be convinced by a demonstration of high-fidelity teleportation in any case.

In principle then Quinten would have to test ``all'' coherent states if he is to use $\tilde{F}=1/2$ as threshold. Since that is impossible he would instead draw randomly from the set of coherent states (varying phase and amplitude randomly) and teleport those and measure the average fidelity. 
On the other hand, small test sets of (nonorthogonal) coherent states may well be sufficient too, provided Quinten beats the correct fidelity limit $\tilde{F}$ (some number larger than 1/2 but less than 1).

In \cite{CVent} it is again Criterion \ref{symm} that was violated by not taking into account the complete set of coherent states that were in fact teleported experimentally for the estimation of the fidelity.
Namely, the fidelity was estimated using the teleportation of one particular coherent state, namely the vacuum. Note that later more general states,
squeezed states in particular \cite{akira2}, were teleported with fidelity $F>1/2$. 
\subsection{Teleportation and hidden variables}
There are two interesting issues concerning the relation between hidden-variable models and
the use of entanglement in teleportation. 

First, there is the question whether a local realistic hidden-variable model exists for the teleportation protocol.
Thanks to the result of \cite{toner} we know now that one bit of classical information suffices to simulate spin-spin correlations of a maximally entangled 2-qubit state. This, as explained in \cite{toner}, can be exploited
by Alice and Bob to mimic a teleportation protocol by making use of the 2 classical bits that Alice is supposed to send to Bob.
However, neither Victor nor Quinten will be fooled by this: Victor not, because he
will check the fidelity of the state on Bob's side himself; Quinten not, because he performs the whole teleportation protocol himself. 

Somewhat similarly, in the case of teleportation with two-mode squeezed states \cite{CVent}, all measurements
(namely, quadrature measurements) and states (Gaussian states only) featuring in that protocol
can be described by a positive Wigner function, which can act as a hidden-variable model \cite{cw}. Again, this is of no concern to Quinten: for him it is sufficient that the only consistent quantum description of the experiment requires entangled states. In particular, his predictions about measurements on the teleported system other than measurements of quadrature variables would be different from those of any hidden-variable model.

Second, it is known that the so-called Werner states $W_\alpha$ \cite{werner} of two qubits
for certain ranges of the parameter $\alpha$ are entangled (for $\alpha>1/3$) but do admit a hidden-variable model for von Neumann measurements on the qubits (for $\alpha\leq 1/2$). Moreover, one can achieve teleportation of qubits with the state $W_\alpha$ for $\alpha=1/2$
and reach a fidelity of $F=3/4$, thus beating the limit of $\tilde{F}=2/3$, as shown in \cite{barrett}.
So in this case, too, Quinten and Victor would conclude there is entanglement in that case, although Rhiannon could find a hidden-variable model.
It may be worth repeating that
even though there is no hidden-variable model according to the criteria of \cite{barrett} when the teleportation fidelity is larger than $F\approx 0.85$ (namely when $\alpha>1/\sqrt{2}$), Rhiannon probably would still not agree with that conclusion as she may exploit the two classical bits sent from Alice to Bob to construct a hidden-variable model, along the lines of \cite{toner}.
\section{Bell and CHSH inequalities}\label{bell}
The underlying asumptions behind Bell and CHSH inequalities have been discussed at length and need no repeating \cite{bell,chsh}.
Nevertheless, we wish here to connect those discussions to the Criteria formulated in Section 2.
In particular, many discussions \cite{santos} center around so-called ``loopholes'': the two most important ones are
the detection (or fair sampling) loophole,
and the locality loophole. Although both loopholes have been closed in separate experiments \cite{loop1,loop2}, there has not been an experiment in which both were closed at the same time \cite{santos}.

The detection loophole concerns the simple fact that in an experiment not all
experimental runs lead to detector clicks, due to inefficiencies in the detectors (losses in the transmission of quantum states from Alice to Bob are part of the generation procedure).
What one assumes is that it is a {\em random} subset of events that is
detected. This assumption does not violate any of the Criteria we proposed in Section 2. Indeed it is an assumption about Quinten's measurement devices, not Alice's and Bob's. Thus Quinten is justified in accepting that assumption, although Rhiannon would not agree. It is assumed here that Alice and Bob do not know which device settings Quinten is going to use, which Quinten must guarantee by choosing his settings randomly.

The locality loophole concerns the timing of the choice of different measurements that
have to be performed on the two systems of the bipartite entangled state. If a Bell inequality test is to refute a {\em local} realistic theory, the two measurements themselves {\em and} choosing the settings of the two measurement devices must be separated by a space-like interval. Most experiments violate that condition but for Quinten's purposes
violating locality is not against the Criteria of Section 2. It is an assumption about {\em his} measurements, not Alice's and Bob's. And so this loophole is not relevant for him, although
it is crucial to Rhiannon. Indeed, Quinten does not even have to make an active choice of settings.
This also relates then to another aspect of testing local realism, namely that of free will of the experimentalist. Rhiannon has to assume 
free will on the experimentalist's part (otherwise the derivation of Bell or CHSH inequalities fails). This implies that an {\em active} choice
of measurements must be made according to Rhiannon. A {\em passive} choice of measurements, for instance determined by a beamsplitter that splits the incoming signal either to one measurement setup or another, is unacceptable for Rhiannon. In contrast, for Quinten both passive and active choices of different measurements are fine, as long as he believes his active choices are permutation invariant (for applying the De Finetti theorem (\ref{simpledF})).

In short, then, closing loopholes is {\em not} important for entanglement verfication, no matter how crucial 
it is for refuting local realistic hidden-variable models.
\subsection{Dangers of postselection}
 In many experiments on polarization entanglement between two photons generated from down conversion, it is common to take into account only those data where two (or more) photons were detected, (at least) one on each side $A$ and $B$, and ignore the null results where photons were not detected on both sides.
For Quinten this is a correct procedure, one reason being it relies only on
the fair-sampling assumption. Another way to see that this procedure is correct, is to note it can be simulated by a local filtering operation. We may imagine on each side independently a polarization-independent lossy channel, 
then a quantum-non-demolition (QND) measurement of photon number, and filtering out those cases where no photons are found.  
Here one would use the relation
\begin{equation}
	E(\rho_{AB})\geq p_{11}E(\rho_{11})+p_{{\rm null}}E(\rho_{{\rm null}}),
\end{equation}
in obvious notation. So postselection correctly identifies the presence of entanglement.
On the other hand, violating a Bell inequality by many standard deviations still does not say much about how much entanglement is generated in the experiment, $E(\rho_{AB})$. If the probability of successful measurements is
small, say $p_{11}=\epsilon$, one may conclude only that the entanglement is of order $\epsilon$, namely $\epsilon E(\rho_{11})$, in contrast to statements in Ref.~\cite{kwiat2}, and in many other papers. The state that may be close to maximally entangled is the fictituous state $\rho_{11}$ that would be produced if one actually performed the above-mentioned QND measurement of photon number.

In other types of experiments the seemingly similar postselection of keeping only data corresponding to detector click fails \cite{kuzmich,kuzmichcomment}. Suppose one intends to generate an entangled state of the form
\begin{equation}\label{0110}
(	|0\rangle|1\rangle+\exp(i\phi)|1\rangle|0\rangle)/\sqrt{2},
\end{equation}
where now 0 and 1 refer to photon numbers in two different modes, one in location $A$, one in $B$.
 Measurements on that state will not always yield one photon in total due to inefficiencies.
 So, why not just ignore the zero-detection results? 
 One reason is that filtering states with 1 photon {\em in total} is a {\em nonlocal} filter, unlike filtering 1 photon {\em on each side}. Indeed, if one instead had generated a product state of the form
 \begin{equation}\label{product}
(	|0\rangle+\epsilon|1\rangle)(\epsilon\exp(i\phi)|1\rangle+|0\rangle)/2,
\end{equation}
with $\epsilon\ll 1$, filtering out zero-detection events would make this state indistinguishable from
the entangled state (\ref{0110}) (if we also neglect the double-detection events arising from the $\epsilon^2|1\rangle|1\rangle$ term), as was pointed out in \cite{kuzmichcomment}.

Alternatively, one may also view this as a violation of Criterion \ref{state}, as the verification protocol of \cite{kuzmich} assumes a single-photon state was generated, explicitly excluding the $|1\rangle|1\rangle$ and
$|0\rangle|0\rangle$ terms by {\em fiat}. So, no entanglement between two atomic ensembles can be inferred from the data presented in \cite{kuzmich}.
\section{Quantum tomography}\label{tomo}
If Quinten performs tomography on the quantum state generated by Alice and Bob, he obviously will not only determine whether the state is entangled but also by how much. However, in general one does not perform a {\em full} tomographic reconstruction of a state, but instead focuses on the subspace or subsystem of interest.
For example, when testing entanglement between the spin degrees of freedom of two electrons, no one would think of also mapping out the spatial degrees of freedom of the electrons. Fortunately, ignoring one degree of freedom is easy to justify: tracing out a degree of freedom is a local operation and can only decrease the amount of entanglement one estimates.

It is a trickier question whether Quinten is allowed to single out some Hilbert {\em subspace} ${\cal H}$ on which to perform his measurements.
Indeed, the example treated at the end of the preceding Section is one where singling out a particular subspace is incorrect.
Moreover, according to Criterion \ref{state} he should not make any assumptions about the state: How can this Criterion be reconciled
with the choice of a particular Hilbert space? 

The answer is this: {\em if} one can show that the overall entanglement in the state generated by Alice and Bob cannot be less than that present in the subspace ${\cal H}$ then, of course, Quinten's test can only underestimate the amount of entanglement.
The most straightforward way of accomplishing this is to make sure that projecting
onto the subspace ${\cal H}$ is a local filtering operation.
This is indeed a method often used, although the restriction to a particular Hilbert subspace is not always made explicit.  For example, tomography on ``photon pairs generated by down conversion'' is typically restricted to the Hilbert space where the number of photons is fixed to two (one on Alice's side, one of Bob's side) \cite{kwiat}. Down conversion in fact generates a mixture of the vacuum, photons pairs, double photon pairs, etc. For the question whether there is entanglement or not, tomography on the restricted Hilbert space only is indeed sufficient, although for quantitative estimates of entanglement it fails. (Let us also note there {\em are} experiments in which tomography is performed on a larger Hilbert space, including the vacuum component as well as components containing photons \cite{lvovsky}.)

Let us here also expand on the example mentioned in the Introduction.
In \cite{chou} tomography is applied to a subspace of the full Hilbert space of two field modes. (The assumption of only two field modes, one on each side, is warranted as ignoring information about what field mode produced a click in one's detectors is a local operation.)
In that experiment,
there is a phase $\phi$ between the two states $|0\rangle|1\rangle$ and $|1\rangle|0\rangle$ that depends on an optical path length. That path length in the actual experiment was controlled and stabilized. Here, as always, one really means that the phase is stabilized with respect to some external reference, so that one really should write $\phi-\phi_R$.
Now the tomographic measurements depend on a similar phase $\phi'-\phi_R$ determined by a different optical path length. The procedure of \cite{chou} made a point of {\em not} using the same optical path for tomography and entanglement generation, so as not to violate Criterion \ref{independence}.

Indeed, suppose that both phases would be drifting over time. Then one could eliminate the relative phase drift of $(\phi-\phi_R)-(\phi'-\phi_R)=\phi-\phi'$ by using the {\em same} optical paths, thus reducing the requirements on phase stabilization. That is, although neither $\phi-\phi_R$ would be well-defined, nor $\phi'-\phi_R$, the difference $\phi-\phi'$ would drift over a much longer time scale. However, that procedure would violate Criterion \ref{independence} as the verification would depend on the generation procedure. And this would have led to premature conclusions about entanglement.

Finally, let us note that assuming the De Finetti form (\ref{simpledF}) is a crucial (albeit often not explicitly noted) part of standard quantum tomography, as explained in \cite{finetti}. As a consequence Criterion \ref{multi} is usually obeyed in such experiments.
\section{Entanglement witnesses}\label{ew}
Entanglement witnesses are operators $W$ on  bi-partite  Hilbert 
spaces such that the observation of  Tr$(\rho W)<0$ implies that $\rho$ is 
an entangled state. Conversely, for each entangled state $\rho$ there is a witness.
Any witness operator can be represented as $W= 
\sum_{i,j} c_{i j} F_i \otimes G_j$, where $c_i$ are real numbers, not 
necessarily positive,  and the sets of operators $\{F_i\}$ and $\{G_j\}$ are POVMs on Alice's and Bob's Hilbert spaces, respectively. Then the 
expectation value of the witness can be evaluated from the observed 
probabilities $p_{i j}= {\rm Tr}(\rho F_i \otimes G_j)$ as Tr$(\rho W)=\sum_{i,j}c_{i j} 
p_{i j}$. Thus an entanglement witness can always be measured by local measurements.

Before looking at some examples, let us make some brief general remarks.
The POVMs on either side do not need to form an operator 
basis (i.e., to be tomographically complete) in order to construct useful witness 
operators \cite{curty}. 
In general we assume that Quinten has full knowledge about the  
POVMs he performs. However, this is not necessary. For instance, there is no implication that Alice and Bob (or Quinten) need to 
share a reference frame in order to be able to verify entanglement. Indeed, 
even if Quinten really uses on Bob's side the POVM $\tilde{G_j}= U G_j 
U^{\dagger}$ for some unknown, but fixed unitary operator $U$ acting on Bob's quantum system,  the 
observation that $\sum_{i,j}c_{i j} p_{i j} < 0$ still verifies entanglement. Namely, in that 
case this observation corresponds to Tr$(I \otimes U \rho I 
\otimes U^{\dagger} W)< 0$, so that he verified the entanglement of the 
state $I \otimes U \rho I \otimes U^{\dagger}$. The latter state is manifestly
connected via a local operation to the original state and, therefore,
entanglement of one state implies the entanglement of the other. This observation can be 
generalized to POVMs related by LOCC (local operations plus classical communication)
maps (rather than unitaries), so that indeed only partial control over the measurement POVMs is 
required. 

Moreover, an entanglement witness does not necessarily make use of {\em all} POVM 
elements of the measurement, therefore trivially allowing the use of local 
filtering. This includes conditioning on the detection of photons, as discussed in the preceding Section. 
Known and well-characterized imperfections such as dark counts and detection 
inefficiencies in Quinten's devices can be directly included in the description of the measurement 
POVM. 

In actual experiments it is crucial not to make (explicit or hidden) assumptions about 
the relevant Hilbert space. For example, in quantum optical 
implementations one uses often single-photon avalanche detectors that monitor many 
spatio-temporal modes. These detectors cannot discriminate from which 
mode the photon has been drawn that triggered the detection event, or 
whether the event was triggered by one or many photons. While 
the issue of many spatio-temporal modes can be easily dealt with due to 
the simple overall structure of the POVM (or by a local filtering operation: measuring but subsequently forgetting from which mode a detected photon arose is a local operation), it is in general much harder
to
analyze carefully multi-photon events.

Entanglement witnesses are being evaluated using the joint probability distribution for the measurement outcomes of both subsystems $A$ and $B$. In a way, this joint probability distribution summarizes {\em everything} Quinten knows about the quantum state. That is, his assignment of the form (\ref{dF}) takes into account all correlations he measured. If the data actually arise from Quinten performing some other entanglement verification protocol, then we know that the results can always  be used to evaluate entanglement witnesses as well.  Thus, whenever some protocol tells Quinten there is entanglement, he can also construct a witness from the same data that reveals entanglement.
Especially those verification methods with criteria that are linear in the density matrix can be rephrased directly as an entanglement witness. An obvious example and a not-so-obvious example follow here.  

Conversely, the data obtained in a measurement of an entanglement witness can be used to give a lower bound on the amount of entanglement. Namely, one can search for states with the minimum amount of entanglement (for one's favorite measure of entanglement) consistent with the data. Thus an entanglement witness can be used both as a qualitative test and as a quantitative measurement of entanglement. This is discussed in various recent papers \cite{reimpel,eisert,plenio}.

Finally, note that if there is no witness to verify entanglement, all other verification methods must fail, too. In this sense, entanglement witnesses represent the strongest methods of entanglement verification.

\subsection{Relation to Bell inequalities}
A Bell inequality test can be related to an entanglement witness \cite{ewbell,hyllus}, as the Bell inequality can 
be expressed as the expectation value of a suitable operator. 
When one does that, one finds typically that the witness operator thus constructed is not optimal: there are better witnesses that detect all states detected by the Bell inequality tests, and more.
This again shows that detecting entanglement is an easier task than refuting local realism.

Note that the Bell witness can be evaluated via a POVM description 
of the measurements, which may include a {\em passive} probabilistic 
choise of measurement settings. This is fine for Quinten, but for Rhiannon such a passive 
detection set-up is not sufficient: she necessarily requires the active 
choice of different settings.
\subsection{Relation to teleportation}
It is interesting to make the connection between entanglement witnesses 
and the teleportation Criteria of Section \ref{telep}. To summarize that procedure,
Quinten 
teleports signals that are drawn at random out of a specified set of states 
with a specified {\em a priori} probabilities. Then he performs measurements 
on Bob's site, so that for the sub-ensemble of each signal, he can 
evaluate the fidelity of the teleportated state. By comparison to classical 
limits to the average fidelity for all states, Quinten then concludes that 
the teleportation actually must have made use of a quantum resource, 
which in this case means that the bi-partite state in teleportation must 
have been entangled. A different way of saying the same thing is that Quinten infers from his data 
that the 
quantum channel, which is realized by the teleportation protocol, is not  entanglement breaking. 

 We can rephrase this whole procedure as a special case of entanglement witnesses. 
After all, if Quinten chooses an input state and performs a Bell 
measurement on this chosen state and Alice's half of the possibly 
entangled state, then he performs an effective POVM on Alice's state
(this trick was used in \cite{barrett} to connect teleportation to Bell inequality tests). 
On Bob's side, Quinten also performs some measurement that allows 
him to reconstruct the conditional states so that he can calculate the 
overlap between input and output state. So the basic data material from the teleportation test
can be interpreted as the measurement of some entanglement witness. 
Indeed, comparing the average fidelity with the optimal classical 
fidelity can be formulated as a particular linear witness. Since there is no reason for the teleportation fidelity to be an optimal witness, it generally does mean 
that just calculating the average fidelity might not allow the verification 
of entanglement, while a more general entanglement witness, constructed from the same data, would do. That is, the same data that yield a fidelity below the limit needed to infer entanglement, may be combined in a different way to demonstrate entanglement.

Combining this picture of teleportation as an entanglement witness with the remarks made above about including local filtering operations shows that  {\em conditional} teleportation is allowed as an entanglement verification protocol.
The conditioning can be seen as a local filtering operation.
\section{Direct measurement of entanglement}\label{direct}
All methods for entanglement verification discussed so far are indirect:
they either allow Quinten to infer about the existence of entanglement by detecting some other property that requires entanglement, or they allow him to reconstruct (the relevant part of) a density matrix that in turns allows him to quantify the amount of entanglement. However, there are {\em direct} measurements that measure the entanglement (either quantitatively or qualitatively) without measuring much more. 
Moreover, such measurements allow one to detect all entangled states, in contrast to a fixed entanglement witness, who can detect only certain entangled states. 
Such measurements require {\em multiple} copies of the same density matrix, entanglement being a nonlinear function of $\rho$. 

For example, Quinten may apply the method of \cite{horodecki2} to detect entanglement by performing suitable measurements on four copies of Alice's and Bob's states.
However, it is not sufficient to have Alice and Bob create just four copies. Alice and Bob could cheat then, if they know Quinten is going to apply that particular method, by preparing an unentangled
4-qubit state with appropriate properties with respect to the observables Quinten will measure. The reason that trick works is that Quinten would wrongly assume the form (\ref{simpledF}) for the state of the four copies.
So, as explained in Section \ref{criterion}, what he should do instead is let Alice and Bob create many (i.e. many more than four) copies; then choose randomly groups of four copies (including a random order within the group); and perform his various measurements on those groups of four in some random order. Finally, he should check that deleting random subsets of data do not change his results.
All this does imply the direct measurement is, in practice, not as good a method as it may seem: one needs to generate many copies at the same time, whereas, e.g., tomography can be done on sequentially produced copies of the (entangled) state. 

The method of \cite{horodecki2} tells Quinten whether there is entanglement or not, but the method does not quantify it.
For that (more lofty) purpose, Quinten can apply the method discussed in \cite{horodecki}: he will have to perform measurements on twenty copies, and estimate the four eigenvalues $\lambda_1\ldots\lambda_4$ (in nonincreasing order) of the nonhermitian matrix $\rho\tilde{\rho}$, where 
\begin{equation}
\tilde{\rho}=\sigma_y\otimes\sigma_y \rho^* \sigma_y\otimes\sigma_y,
\end{equation}
where $\rho^*$ denotes the complex conjugate of $\rho$ in the standard basis.
The eigenvalues together determine the concurrence of one copy of Alice's and Bob's state through \cite{wootters}
\begin{equation}
C(\rho)=\max(0,\sqrt{\lambda_1}-\sqrt{\lambda_2}-\sqrt{\lambda_3}-\sqrt{\lambda_4}).
\end{equation}
But just as in the preceding example, Quinten does have to be careful as the method of \cite{horodecki} relies on all twenty copies being in the same state.
Just as before, he should take a large number of copies generated by Alice and Bob, and choose random subsets of twenty copies, and then perform the measurements of \cite{horodecki} in some random order.
That will give him a correct estimate of the amount of entanglement, provided, again, the random deletion of subsets of data does not change his quantitative estimate of entanglement (within experimental error, of course). 

A possible improvement over the method  of \cite{horodecki} is described in \cite{carteret}, in which circuits are constructed for measuring the concurrence directly, again using multiple copies.
We note that those circuits are not complete and must be supplemented by random permutations of the input states in order to turn the measurement into a valid entanglement-verification protocol.

Finally, let us discuss an example of how violating the Criteria leads one to overestimate entanglement even with the direct measurement methods. 
In Ref.~\cite{walborn} an experiment is described where the concurrence of an entangled state is estimated using a direct measurement of entanglement in
a particularly simple way. The method relies on having two copies of the {\em same pure state}.
This is obviously a much stronger condition than assuming the state to be of the form (\ref{simpledF}). Indeed, if Quinten would assume the general De Finetti form including mixed states---and Quinten is never allowed to assume more than that--- the method of \cite{walborn} fails (for details see \cite{enkcom}).  We also note that, according to \cite{rbk},
a finite set of measurements never should lead to a pure-state assignment. An indication that the procedure of \cite{walborn} relies on an overtly strong assumption is that all measurements
take place entirely on Alice's side.

More precisely, the method of \cite{walborn} is based on the following (correct) theoretical result \cite{mintert}: {\em if} two bipartite states are {\em pure} and identical, then the concurrence
of a single copy of the state is given by
\begin{equation}\label{C}
C=2\sqrt{P_a^A},	
\end{equation}
where $P_a^A$ denotes the probability to project the two qubits on Alice's side onto the antisymmetric subspace. The experiment now assumes the source generates two identical copies, measures $P_a^A$ on Alice's particle only, and concludes that maximal entanglement has been generated from the observation that $P_a^A=1/4$.
However, if Alice creates the completely mixed state $\rho_A=I/2$ on her side instead, one will also find that $P_a^A=1/4$. 

A further problem is that in the experimental setup of Ref.~\cite{walborn} the joint measurements can only be done on specific pairs of systems.
It is clear, in that case, that Alice and Bob can fool a Quinten using this method into believing they share maximal entanglement by yet another procedure: Alice can create pairs of particles, 1/4 of them in the singlet state, the remaining 3/4 can be in randomly chosen product states $|0\rangle|0\rangle$ or $|1\rangle|1\rangle$. 
The improved method \cite{flo} of measuring $P_a$ both on Alice's and Bob's side and verifying the correlations between these two measurements would succumb to the same cheating method in the experiment. Whenever Alice creates a singlet state on her side, Bob does the same on his side. 

Unfortunately, a very recent paper \cite{new} 
follows essentially the same argumentation of \cite{walborn}, by proposing a direct measurement of entanglement on just Alice's system that, again, relies on assuming {\em pure} states. Obviously, measurements on Alice's system only, no matter what observables one measures,
can never tell one anything about entanglement, as the maximally entangled state is then indistinguishable from the completely mixed state. All these proposed methods are simple only because the missing part of the proof of entanglement, namely that
 one's source produces identical pure states, is the most difficult.

This version of the direct measurement method is somewhat similar to the (incorrect) protocol 6 we mentioned in the Introduction, {\em consistency with entanglement}.
To illustrate this with a simple example, suppose we assume we create a pure entangled state of the form
\begin{equation}\label{knudde}
\sin\theta |0\rangle|1\rangle+\cos\theta |1\rangle|0\rangle,
\end{equation}
where $\theta$ is a control parameter.
Then we might think we directly measure entanglement if we just
estimate the probability of finding system $A$ in state $|0\rangle$ or $|1\rangle$.
After all, that measurement determines $\sin^2\theta$ and thereby the entanglement in the state (\ref{knudde}). But, it should be clear this measurement in fact only checks for consistency with the state assignment (\ref{knudde}) without verifying or demonstrating entanglement. Demonstrating entanglement would require one to verify the form (\ref{knudde}). 
\section{Discussion}
We have now discussed many different ways to verify entanglement.
We distinguished between protocols that eliminate all possible local hidden-variable models from 
those protocols that accept quantum mechanics as a valid description of Nature
and infer entanglement. We argued that the latter protocols are sufficient for entanglement verification.
In particular, an important distinction is that closing loopholes is {\em not} important for entanglement verfication, no matter how crucial 
it is for refuting local realistic hidden-variable models.

On the other hand, we argued that an entanglement verification protocol
should not put any trust in the entanglement generation procedure.
After all, if one would completely trust the generation procedure there would be no need for entanglement verification. In order to set a sharp boundary 
we proposed to pretend the entangled states one tries to verify were handed over by untrusted parties.

We discussed three different flavors of entanglement.
The distinction we made between the three different types is relevant for our discussions, as confusing one type for another has often lead to
incorrect interpretations of the results of entanglement verification protocols.

We discussed a number of different entanglement-verification protocols: teleportation, violating Bell-CHSH inequalities, quantum state tomography, entanglement witnesses, and direct measurements of entanglement. Let us give an interpretation of some of these procedures.  Generally speaking, a successful entanglement verification protocol teaches one the following: if one's source produced many
instances of the state that was tested and verified, then the remaining untested states are guaranteed to possess entanglement. This is true provided the remaining states form a {\em random} subset of all states generated by the source. However, in all of the entanglement-verification protocols we discussed, except {\em full} quantum-state tomography, the state itself is still not completely known. 
This is true for direct measurements, it is true for entanglement witnesses, and it is true for Bell-CHSH inequality tests. For the latter, it in fact tends to be a great advantage that all one needs is strong correlations between certain measurement outcomes, without knowing what actually was measured, and without knowing much about the Hilbert space structure of the system. But it does imply one will not know what quantum state to assign to the remaining untested copies. 

All this means in particular that the remaining untested states may not be used yet for, e.g., teleportation.
Similarly, although certain entanglement verification protocols allow one to use the
untested states for quantum key distribution---measuring entanglement witnesses of an appropriate form in particular--- others do not. Thus, if one wishes to use entanglement for teleportation, then one could certainly first perform, say, the direct measurements discussed in Section \ref{direct} to determine first if there is in fact entanglement. However, one still has to follow up with more refined measurements, e.g., tomographic measurements, to narrow down one's possible state assignment. This does lose some of the advantage of the direct measurement methods.
Both quantum key distribution and teleportation as entanglement verification protocols have the advantage, in this context, that the protocol itself performs some useful task while verifiying the presence of entanglement at the same time. 

Finally, we identified many pitfalls associated with entanglement verification and quantification.
We formulated five Criteria that, we think, should be applied to any experimental entanglement verification protocol. This, we hope, will help in unifying the language used for describing the different types of entanglement that can be created in a large variety of physical systems. That should also make it easier to compare in a consistent fashion and operationally useful way the various types of entanglement created in experiments.
\section*{Acknowledgments}
We thank R. Blume-Kohout and M. Raymer for discussions.
The research of NL has been funded by the European Union through the IST-FET Integrated Project QAP.
The research of HJK is supported by the National Science Foundation and by
the Disruptive Technologies Office (DTO) of the DNI. 


\begin{thebibliography}{99}
\bibitem{werner}R.~Werner, Phys. Rev. A {\bf 40}, 4277 (1989);
The Werner state $W_\alpha$ is
of the form $\alpha P_s+(1-\alpha)I/4$, where $P_s$ is the projector onto the single state
and $I$ is the identity.
\end{thebibliography}
\end{document}